\documentclass[12pt]{revtex4-2}      
\usepackage{graphicx}
\usepackage{amssymb}
\usepackage{amsmath}

\usepackage{color}
\newcommand{\red}[1]{{#1}}
\newcommand{\eff}{\mathrm{eff}}

\newcommand{\GeV}{\mathrm{GeV}}
\newcommand{\br}{\mathrm{Br}}
\begin{document}
\title{The factorization effect in $\Lambda_C\to \Lambda\pi$ and $\Lambda_C\to \Lambda K$ }
\author{Shou-shan Bao}
\email{ssbao@sdu.edu.cn}
\affiliation{Institute of Frontier and Interdisciplinary Science,\\
Key Laboratory of Particle Physics and Particle Irradiation, Shandong University, Qingdao 266237,P. R. China}
\begin{abstract}
We calculated the branch ratios of $\Lambda_C\to \Lambda K$ and $\Lambda_C\to \Lambda \pi$ within factorization approximation. It is found that the results suffers from the uncertainties of form factors. To reduce such uncertainties, we checked the ratio of the two channels, which is obtained as $R=0.0743\pm 0.0014$. It is found that the uncertainty caused by form factors is significantly reduced. This note is prepared for the BESIII collaboration.
\end{abstract}
\maketitle
\section{Introduction}
Although the standard model (SM) has achieved great success, there are remaining some problems with no answers. Precision studies of the hadron decay processes through weak interaction play an important role in testing  SM and indirect searching new physics.  During recent years, tremendous progress in bottom physics and charm physics has been made through the fruitful interplay between theory and experiment. The precise measurements of the weak decay channels can provide an insight into high energy scales new physics. 

The factorization approximation has been applied successfully in B-physics. Due to the low scale, it is thought  that charm physics would suffer from large non-perturbation effect and the un-negligible non-factorization contribution~\cite{Ke:2020uks,Geng:2017esc,Chen:2002jr,Cheng:2018hwl,Sharma:1996sc,Uppal:1994pt}.  Numerically, the factorization contributions are still crucial in $\Lambda_C\to \Lambda\pi$ and  $\Lambda_C\to \Lambda K$. We perform a more careful calculation with the form factors from Ref.~\cite{Meinel:2016dqj}, and show the uncertainties.

There are many works about the transition $\Lambda_C\to \Lambda$ have been done, and some of that can be found in Ref.\cite{Cheng:2018hwl} and its references. We re-visit the topic because large data-sets  are available at the BESIII experiment which would get better understanding on the charmed baryons. In this work we calculated the factorization contributions to $\Lambda_C\to \Lambda+\pi/K$.  With the form factors from Ref.{\cite{Meinel:2016dqj}}, we get the following results.
\begin{align}
&\br(\Lambda\, \pi)=(1.41\pm 0.15) \times 10^{-2}, \quad \alpha=0.992\pm 0.013,\\
&\br(\Lambda K)=(1.05\pm 0.10) \times 10^{-3}, \quad \alpha=0.982\pm 0.017.
\end{align}
Only the uncertainties of the form factors are considered in the results.  One can see the results suffer from large uncertainties. To reduce the uncertainties caused by form factors, the ratio of the two branch ratio is obtained.
\begin{align}
R=\frac{\br(\Lambda K)}{\br(\Lambda \pi)}=0.0743\pm 0.0014.
\end{align}
One can find the uncertainty of the ratio given in factorization approximation is small. 

This paper is organized as follows: In Sec. II, we present the general formulas for the decay of $\Lambda_C\to \Lambda P$. In Sec.III, we list the inputs and present the numerical results. Finally, the conclusion is presented in Sec. IV.
\section{The general formulas}
The $\Lambda_C\to \Lambda $ decay is induced by the effective Hamiltonian as follows\cite{Buchalla:1995vs}
\begin{align}
\mathcal{H}_{\eff}= \frac{G_F}{\sqrt{2}}\left( \sum_{q=d,s} V^*_{cq} V_{uq} ( C_1 Q^q_1+C_2 Q^q_2)+ V^*_{cb} V_{ub} \sum_{i=3,6}C_i Q_i\right).
\end{align}
The effective operators are as follows.
\begin{align}
Q_{1,q}&=(\bar{s}^\alpha c^\beta)_{V-A} (\bar{u}^\beta q^\alpha)_{V-A},\\
Q_{2,q}&=(\bar{s}^\alpha c^\alpha)_{V-A} (\bar{u}^\beta  q^\beta )_{V-A},\\
Q_{3,5}&=(\bar{u}^\alpha c^\alpha)_{V-A} \sum_q (\bar{q}^\beta q^\beta )_{V\mp A},\\
Q_{4,6}&=(\bar{u}^\alpha c^\beta )_{V-A} \sum_q (\bar{q}^\beta  q^\alpha)_{V\mp A}.
\end{align}
where the $(\alpha,\beta)$ are the color indices. 
The numerical values of Wilson coefficients and their running with scale can be found in Ref.{\cite{Li:2012cfa}}.


For $\Lambda_C\to \Lambda \pi$, only the tree-operators $Q_1$ and $Q_2$ have contributions. And for $\Lambda_C\to \Lambda K$, the contribution of  $Q_1$ and $Q_2$ are also dominated. In the factorization approximation, the transition amplitude is factorized into the product of two matrix elements of single currents, i.e. $\langle \Lambda \vert (V-A)\vert \Lambda_C\rangle \times \langle P\vert (V-A) \vert0 \rangle$. Here we use $V(A)$ to denote the (axial-)vector current operators. The hadron transition matrix elements of $\Lambda_C\to \Lambda$ can be parametrized with six from factors\cite{Weinberg:1958ut,Detmold:2016pkz}. 
\begin{align}
\langle \Lambda(p^\prime)\vert V^\mu \vert \Lambda_c(p)\rangle&=\bar{u}_{\Lambda} \left(\gamma^\mu f_1 \red{+} i q_\nu \sigma^{\mu\nu} f_2+ q^\mu f_3\right)u_{\Lambda_c}(p),\label{eq:hmea}\\
\langle \Lambda(p^\prime)\vert A^\mu \vert \Lambda_c(p)\rangle&=\bar{u}_{\Lambda} \left(\gamma^\mu g_1\red{+} i q_\nu \sigma^{\mu\nu} g_2+ q^\mu g_3\right) \gamma^5u_{\Lambda_c}(p).\label{eq:hmev}
\end{align}
The numerical values of form factors can be found in Ref.\cite{Meinel:2016dqj}.

The contributions from the operators $Q_{3}$ and $Q_4$ can be directly related to that from the tree operators by replacing the Wilson coefficients. The contribution of $(V-A)(V+A)$ current can be transpose to  $(S-P)(S+P)$ current with Fierz identities\cite{Nieves:2003in}. The relation of the contribution from $(S-P)(S+P)$ current and that from the $(V-A)(V-A)$ current is a little subtler and is note listed here. However the total contributions from the penguin operators are negligible. 

The transition amplitude can be expressed as
\begin{align}
\mathcal{M}(\Lambda P)&= \langle \Lambda, P\vert \mathcal{H}_\mathrm{eff}\vert \Lambda_c\rangle= i\bar{u}_\Lambda (A-B\gamma^5) u_{\Lambda_C},\quad P=\pi, K.\\
A&=\frac{G_F}{\sqrt{2}}V_{\mathrm{CKM}}C_{\eff} f_P \left(f_1(m_{\Lambda_c}-m_\Lambda)+f_3 m_P^2\right),\label{eq:a}\\
B&=\frac{G_F}{\sqrt{2}}V_{\mathrm{CKM}}C_{\eff} f_P \left(-g_1(m_{\Lambda_c}+m_\Lambda)+g_3 m_P^2\right),\label{eq:b}
\end{align}
where the $V_{\mathrm{CKM}}$ is the product of two CKM matrix elements. The decay width of $\Lambda_C\to\Lambda \pi/K$ is obtained as
\begin{align}
&\Gamma(\Lambda_C\to \Lambda P)=\frac{q}{8\pi} \frac{(m_{\Lambda_C}+m_{\Lambda})^2-m^2_P}{m_{\Lambda_C}^2} \left(\vert s\vert^2+\vert p\vert^2\right),\label{eq:width}\\
&\qquad q=\frac{1}{2m_{\Lambda_C}}\left({\left(m_{\Lambda_C}^2-m_{\Lambda}^2-m^2_P\right)^2-4 m_{\Lambda}^2 m_P^2}\right)^{{1}/{2}},\\
&\qquad s=A, \quad p=\kappa B, \quad \kappa=\sqrt{\frac{E_{\Lambda}-m_{\Lambda}}{E_{\Lambda}+m_{\Lambda}}}.
\end{align}
The up-down asymmetry is expressed as follows, which can be measured through the polarization of $\Lambda$ in final state.
\begin{align}
\alpha=\frac{2\mathrm{Re}(s^* p)}{(\vert s\vert^2+\vert p\vert^2)}=\frac{2\mathrm{Re}(\kappa A^* B)}{\vert  A\vert^2+\vert\kappa B\vert^2}.
\end{align}
\section{Results and discussion}

\begin{table}[htb]
\centering
\begin{tabular}{ccccccccc}
\hline
$\Gamma_{\Lambda_C}$&$m_{\Lambda_C}$ & $m_\Lambda$ & $m_\pi$ &$f_\pi$  & $m_K$ & $f_K$\\
\hline
$3.25783\times 10^{-12}$&2.28646& 1.115683& 0.13957039&0.1302  & 0.493667 & 0.1557\\
\hline
\end{tabular}
\caption{The values of the input parameters taken from PDG\cite{Zyla:2020zbs}. The unit used in this work is GeV.}\label{tb:inputs}
\end{table}
Some numerical values of the input parameters are listed in Table.(\ref{tb:inputs}). For the CKM metric elements, we use the Wolfenstein parameters as inputs, which are
$ \lambda = 0.22650$, $A = 0.790$,  $\bar{\rho} = 0.141$ and $\bar{\eta} = 0.357$. The Wilson coefficients at the scale $m_C\simeq 1.3\GeV$ are listed in Table.(\ref{tb:wc}).
The parameters and formulas of form factors are not listed here, but can be found in Ref.\cite{Meinel:2016dqj}. The ``nominal fit'' results are used in this work. In Ref.\cite{Meinel:2016dqj}, the parameterizations of the hadron matrix elements are different from Eq.(\ref{eq:hmev}) and Eq.(\ref{eq:hmea}) and their relations can be found in Ref.\cite{Detmold:2016pkz}.
\begin{table}[!htb]
\centering
\begin{tabular}{cccccc}
\hline
$C_1$& $C_2$ & $C_3$ &$ C_4$& $C_5$ &$C_6$\\
\hline
-0.43& 1.22& -0.018 & 0.046 & -0.013 & 0.044\\
\hline
\end{tabular}
\caption{The Wilson coefficients at the scale $m_C\simeq 1.3\GeV$\cite{Li:2012cfa}.}\label{tb:wc}
\end{table}

Finally, the numerical results obtained are
\begin{align}
&\br(\Lambda\, \pi)=(1.41\pm 0.15) \times 10^{-2}, \quad \alpha=0.992\pm 0.013,\\
&\br(\Lambda K)=(1.05\pm 0.10) \times 10^{-3}, \quad \alpha=0.982\pm 0.017.
\end{align}
The uncertainties are caused by the form factors. To reduce the uncertainties, we focus on the ratio of the branch ratio. According to the Eq.(\ref{eq:a}),Eq.(\ref{eq:b}) and Eq.(\ref{eq:width}), if the mass difference of pion and kaon is omitted, the ratio can be found as
\begin{align}
R=\frac{\br(\Lambda K)}{\br(\Lambda \pi)}\to \left\vert\frac{V_{us} f_K}{V_{ud} f_\pi}\right\vert^2=0.076.
\end{align} 
At such limit, the uncertainties of form factors are completely canceled. When the mass difference taken into account, we obtain the factorization result as 
\begin{align}
R=\frac{\br(\Lambda K)}{\br(\Lambda \pi)}=0.0743\pm 0.0014.
\end{align}
We can see that the uncertainty from the form factors is largely reduced. The Belle collaboration and BaBar collaboration have presented their results. The factorzation result seems more consistent with the Belle result $R=(7.4\pm1\pm1.2)\%$~\cite{Belle:2001hyr}, the uncertainty of which is larger. However, The BaBar result $R=(4.4\pm0.4\pm0.3)\%$~\cite{BaBar:2006eah} indicates large contribution from non-factorization effect. With large number of $\Lambda_C$ events collected, the BESiII collaboration can give a third measurement and settle down this controversy, which is helpful to clarify the role played by the non-factorization effect in $\Lambda_C$ decays.

\section{Conclusion}
To clarify the non-factorzation effect in $\Lambda_C\to \Lambda$ decays, we need to know the factorization effect more preciously. 
In this work we calculate the branch ratios of $\Lambda_C\to \Lambda \pi/K$ in factorization approximation, and find that the branch ratios suffer from large uncertainties of form factors. To reduce such uncertainties, we calculated the ratio between the two channels.  At the limit of $m_K-m_\pi\to 0$, the form factors effects can be completely removed from the ratio. When we consider the mass difference between the pion and kaon, we obtain the numerical result as $R=0.0743\pm 0.0014$,  in which the effect of form factors is reduced significantly. The BESIII experiment can measure this ratio and clarify the non-factorzation effect in $\Lambda_C$ decays.
\begin{acknowledgments}
The author thanks Z. Q. Liu for useful discussion. This work is supported in part by the Natural Science Foundation of Shandong Province under grant No. ZR2020MA094.
\end{acknowledgments}

\end{document}